\documentclass[sigconf]{acmart}

\copyrightyear{2026}
\acmYear{2026}
\setcopyright{cc}
\setcctype{by}
\acmConference[MSR '26]{23rd International Conference on Mining Software Repositories}{April 13--14, 2026}{Rio de Janeiro, Brazil}
\acmBooktitle{23rd International Conference on Mining Software Repositories (MSR '26), April 13--14, 2026, Rio de Janeiro, Brazil}
\acmPrice{}
\acmDOI{10.1145/3793302.3793304}
\acmISBN{979-8-4007-2474-9/2026/04}

\usepackage{tabularx,booktabs,threeparttable,array}
\usepackage{pifont} 
\usepackage{pifont}                     
\usepackage{multicol,multirow}
\usepackage[table]{xcolor}
\usepackage{lstlinebgrd}
\usepackage{graphicx}
\newcolumntype{W}{>{\raggedright\arraybackslash}X}   
\newcolumntype{C}{>{\centering\arraybackslash}X}     

\usepackage{listings}
\usepackage{xcolor} 
\definecolor{codegreen}{rgb}{0,0.6,0}
\definecolor{codegray}{rgb}{0.5,0.5,0.5}
\definecolor{codepurple}{rgb}{0.58,0.3,0.82}
\definecolor{backcolour}{rgb}{0.95,0.95,0.92}

\lstdefinestyle{mystyle}{
    language=python,
    backgroundcolor=\color{backcolour},   
    commentstyle=\color{codegreen},
    keywordstyle=\color{magenta},
    numberstyle=\tiny\color{codegray},
    stringstyle=\color{codepurple},
    basicstyle=\ttfamily,
    breaklines=true,
    captionpos=b,                    
    keepspaces=true,                 
    numbers=left,                    
    numbersep=5pt,                  
    showspaces=false,                
    showstringspaces=false,
    showtabs=false,                  
    tabsize=2,
    frame=tb,                 
    rulecolor=\color{gray},   
    framerule=0.9pt,
    escapeinside={(*}{*)}
}

\usepackage[most]{tcolorbox}
\usepackage[linesnumbered,ruled,vlined]{algorithm2e}

\definecolor{headbg}{gray}{0.93}

\definecolor{codegreen}{rgb}{0,0.6,0}
\definecolor{codegray}{rgb}{0.5,0.5,0.5}
\definecolor{codepurple}{rgb}{0.58,0.3,0.82}
\definecolor{backcolour}{rgb}{0.95,0.95,0.92}
\definecolor{headbg}{gray}{0.93}

\hypersetup{
    linkcolor=Peach, 
    citecolor=Peach, 
    filecolor=magenta, 
    urlcolor=cyan,
    pdfborderstyle={/S/U/W 0.5}
}

\begin{document}

\title{GLiSE: A Prompt-Driven and ML-Powered Tool for Automated Grey Literature Extraction in Software Engineering}



\author{Brahim MAHMOUDI}
\orcid{0009-0008-3007-7080}
\affiliation{%
  \institution{École de technologie supérieure}
  \city{Montréal}
  \state{Québec}
  \country{Canada}
}
\authornotemark[1]

\author{Zacharie CHENAIL-LARCHER}
\orcid{0009-0002-7464-6659}
\affiliation{%
  \institution{École de technologie supérieure}
  \city{Montréal}
  \state{Québec}
  \country{Canada}
}
\authornotemark[1]

\author{Houcine Abdelkader CHERIEF}
\authornote{The first three authors contributed equally to this work and share first authorship.}
\affiliation{%
  \institution{École de technologie supérieure}
  \city{Montréal}
  \state{Québec}
  \country{Canada}
}
\orcid{0009-0007-4347-1872}

\author{Quentin STIÉVENART}
\orcid{0000-0001-9985-9808}
\affiliation{%
  \institution{Université du Québec à Montréal}
  \city{Montréal}
  \state{Québec}
  \country{Canada}
}

\author{Naouel MOHA}
\orcid{0000-0001-9252-9937}
\affiliation{%
  \institution{École de technologie supérieure}
  \city{Montréal}
  \state{Québec}
  \country{Canada}
}


\author{Florent AVELLANEDA}
\orcid{0000-0003-1030-5388}
\affiliation{%
  \institution{Université du Québec à Montréal}
  \city{Montréal}
  \state{Québec}
  \country{Canada}
}



\begin{abstract}
Grey literature is essential to software engineering research as it captures practices and decisions that rarely appear in academic venues. However, collecting and assessing it at scale remains difficult because of their heterogeneous sources, formats, and APIs that impede reproducible, large-scale synthesis. To address this issue, we present \textsc{GLiSE}, a prompt-driven tool that turns a research topic prompt into platform-specific queries, gathers results from common software-engineering web sources (GitHub, Stack Overflow) and Google Search, and uses embedding-based semantic classifiers to filter and rank results according to their relevance. \textsc{GLiSE} is designed for reproducibility with all settings being configuration-based, and every generated query being accessible.
In this paper, (i) we present the \textsc{GLiSE} tool, (ii) provide a curated dataset of software engineering grey-literature search results classified by semantic relevance to their originating search intent, and (iii) conduct an empirical study on the usability of our tool.
\end{abstract}



\renewcommand{\shortauthors}{Mahmoudi, Chenail-Larcher, Cherief et al.}
\maketitle

\section{Introduction}

Grey literature plays a pivotal role in software engineering research, encompassing technical artifacts disseminated across various sources such as blogs, GitHub issues, official documentation, Q\&A forums, and collaborative development platforms. Unlike curated academic venues, these heterogeneous sources capture the evolving practices surrounding languages, frameworks, and tools, often documenting failures, workarounds, and community norms in real time \cite{garousi2021grey,alves2020grey}. 
Leveraging such evidence can enhance ecological validity and responsiveness to industrial trends, provided that it can be systematically collected, normalized, and assessed at scale\cite{Garousi2020}.

Despite this potential, existing approaches to grey-literature extraction in software engineering remain predominantly manual, relying on ad-hoc scripts or generic search engines \cite{yasin2020using,zhang2022code, shao2024llms}. Repeatability and coverage are limited because search strategies are rarely standardized, collection pipelines are not reusable, and validation procedures are under-specified. In contrast to academic digital libraries, there is still no dedicated framework enabling automated discovery, acquisition, and curation of grey literature at scale\cite{saleh2014grey}.

Several practical barriers further exacerbate this gap. Sources exhibit heterogeneous formats (Markdown pages, threaded discussions, issue-tracking schemas), sparse and inconsistent metadata, and variable quality, complicating de-duplication, and provenance tracking. Together, these factors hinder the construction of representative corpora that can support evidence-based synthesis.

This paper addresses these limitations by introducing \textsc{\textbf{GLiSE}} \cite{GLiSE2025}, an automatic end-to-end tool for the extraction and curation of grey literature in software engineering (SE). The design integrates web-scale search with platform-specific connectors, and natural-language-processing components for classification and ranking. 


The contributions of this work are threefold:
(i) \textbf{\textsc{GLiSE}}, a tool for automatic SE-specific grey literature search, filtering and extraction  \cite{GLiSE2025};
(ii) A manually curated dataset of 1,137 search results and search intent pairs classified by relevance;
(iii) An empirical usability evaluation of our tool.


\section{Related Work}

The systematic use of grey literature in software engineering (SE) has gained traction as a complement to academic sources. Garousi et al.\cite{garousi2021grey} reviewed motivations, benefits, and challenges of integrating non-academic evidence into secondary studies, while Alves et al. \cite{alves2020grey} mapped grey literature types (e.g., blogs, Q\&A, technical docs) and their value for capturing emerging practices. Yet, most studies remain descriptive and rely on manual or semi-manual searches for their extraction, which limit reproducibility and scalability \cite{kitchenham2007guidelines,KALIBATIENE2026104073}.
Evidence-based SE calls for systematic, traceable, and automated pipelines for evidence collection and synthesis \cite{garousi2016evidence,anchundia2020resources}. While recent work improves rigor and present partial automation of data-extraction workflows \cite{cruz2019replication,kitchenham2007guidelines}, few approaches target grey literature explicitly. Schmidt et al. \cite{schmidt2024automated} leveraged NLP and mining techniques to regroup unstructured practitioner artefacts. However, the sources studied (clinical reports, policy briefs, patient-oriented docs) differ substantially from SE artefacts, such as GitHub issues and repositories, Stack Overflow discussions, and API docs. As a result, query design, data representation, and validation strategies are not directly transferable. Effective treatment of SE sources therefore requires dedicated query planning and request handling that account for platform-specific affordances.

While AI-powered search tools such as \textit{Perplexity} \cite{PerplexityAI2025} and chatbots (e.g., ChatGPT) have greatly facilitated general web search, they possess important limitations that make them unsuited for systematic grey literature search. Their retrieval scope is broad rather than source-targeted, they offer very limited transparency and traceability, they provide minimal reproducibility, and they give users little control over search configurations. They are also constrained by context limits, which restrict how many results can be examined at once. By contrast, \textsc{GLiSE} is specifically designed for grey literature search in SE. It targets selected sources, exposes each step of the retrieval process, allows export and import of queries and results for reproducibility, handles larger result sets, and gives users direct control over search options.

\section{Tool Architecture and Workflow}

\begin{figure*}[t]
  \centering  \includegraphics[width=\textwidth,height=.18\textheight,keepaspectratio]{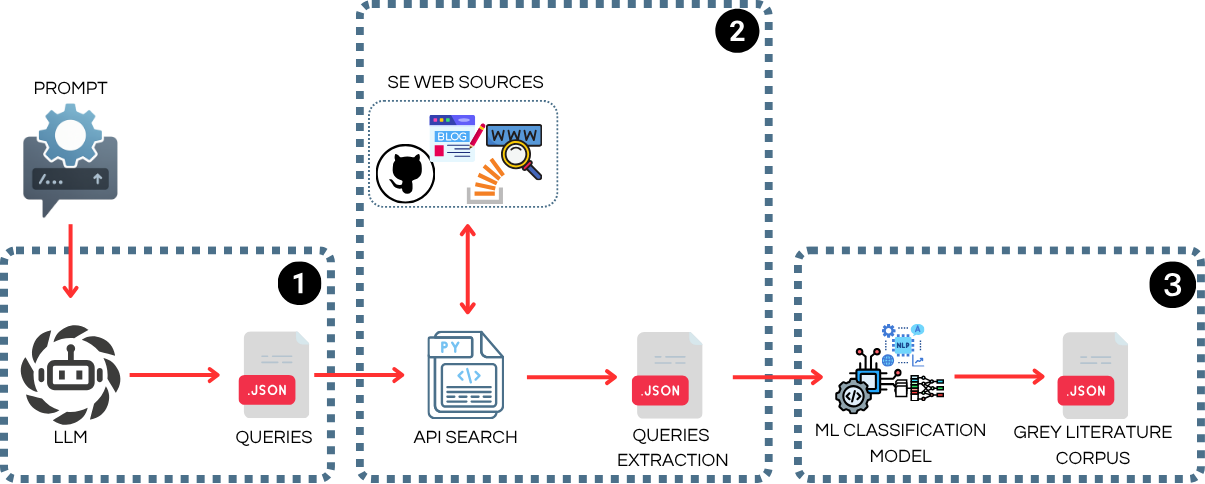}
  \caption{\textsc{GLiSE} prompt-to-corpus workflow}
  \label{fig:grey-literature}
  \Description{In the first step, the LLM takes a prompt to generate queries. In the second step, the search engine API uses these queries to collect results from web sources and generate a “queries extraction” file. In the third step, this file is used by the ML classification model to generate a grey literature corpus}
\end{figure*}

GLiSE'S workflow consists of 3 steps that transform a user-provided prompt into a filtered and ranked corpus of grey literature items as illustrated by figure \ref{fig:grey-literature}. For each step, we present the inputs and outputs, and give a short description of the processing performed.

\subsection{Step~1: Query extraction}

\noindent\textbf{Input:} A free‑text prompt describing the search intent.

\noindent\textbf{Output:} Source-specific search queries.

\noindent\textbf{Description:} This phase performs query planning. From the prompt and the options selected in the interface, \textsc{GLiSE} uses a Large Language Model (LLM) accessed through OpenAI's API \cite{OpenAI_API2025, OpenAI_ModelsDocs2025} to generate platform-specific search queries. Users can set a time range, pick the sources, select the LLM and its temperature, restrict languages, and choose the total number of queries to create. For GitHub, the planner targets repository descriptions, README files, and issues with native qualifiers such as \texttt{in:description}, \texttt{in:readme}, and \texttt{is:issue} together with optional language and date filters. For Stack Overflow, it generates title and body queries with tag constraints and optional signals such as accepted answer or minimum score. For Google Search, it composes expressions with \texttt{site} and \texttt{filetype} operators and synonyms of the topic.
Generated queries can be exported and imported to improve reproducibility.

\subsection{Step~2: API calls}

\noindent\textbf{Input:} Source-specific search queries from Step~1. 

\noindent\textbf{Output:} Search results for each selected source.

\noindent\textbf{Description:} We use the queries from Step 1 and call the public APIs of selected web sources. It paginates results, handles retries, and records provenance for each request. For every result, it extracts core metadata such as URL, title, and snippet. It also extracts source specific data such as \textit{Readme files} for GitHub repositories and \textit{meta descriptions} for Google. The step performs near-duplicate detection over URL, titles and snippets.




\subsection{Step~3: Relevance classification}

\noindent\textbf{Input:} Search results from Step~2.

\noindent\textbf{Output:} Search results filtrated and ranked by relevance.

\noindent\textbf{Description:} For each retrieved item, we compute embeddings (vector encoding of text semantics) for all relevant textual fields (e.g., title, snippet, README) using either \textit{text-embedding-3-small} or \textit{text-embedding-3-large}, depending on the user's indications. We also embed the search intent itself. The embedding of each field is then compared with the search-intent embedding using vector-space operations (e.g. cosine distance, element-wise absolute difference), producing feature values that characterize their semantic similarity.

These features are given to a machine-learning classifier specific to both the data source and the chosen embedding model (for a total of eight). It predicts whether the item is relevant to the search intent. Retrieved items are then ranked based on the classifier's predicted probability of relevance when available, or otherwise by its confidence score.

\section{Data and Model Construction}
\label{sec:precision}

A central function of \textsc{GLiSE} is the filtering of search results according to their relevance to the user's search intent. As introduced earlier, this capability relies on embedding representations of the search intent and of the metadata of the retrieved items, combined with a machine-learning classifier. To design this component, we selected an appropriate embedding model based on practical criteria, and conducted an exploratory study to identify the optimal combination of embedding-derived inputs, embedding dimensionalities, and Machine Learning (ML) models.

\begin{table}[h]
\centering
\setlength{\tabcolsep}{4pt}
\renewcommand{\arraystretch}{0.87}
\caption{Manual labels per source.}
\small
\label{tab:labels}
\begin{tabular}{lrrrrr}
\toprule
\textbf{Sources} & \textbf{N (items)} & \#\textbf{Relevant} & \#\textbf{Irrelevant}  \\
\midrule
GitHub Repos & 222 & 80 & 142  \\
GitHub Issues & 216 & 75 & 141  \\
Stack Overflow & 390 & 269 & 121  \\
Web Search & 309 & 154 & 155 \\
\midrule
\textbf{Total} & 1137 & 570 & 559  \\
\bottomrule
\end{tabular}
\end{table}

\subsection{Dataset Construction}
To conduct our exploratory study and train our models, we first needed to construct an appropriate dataset. We began by defining a set of search intents, which we then transformed into concrete search queries using an LLM, following the same process used within the tool. These generated queries were sent to the APIs of the sources, and the retrieved results were collected and associated to their originating search intent.
Then, for each provider, we manually curated and labeled the retrieved items according to whether they were relevant to their associated search intent or not. This process produced a set suitable for training and evaluating of our filtering models. The resulting dataset is is available in the \textsc{GLiSE} repository~\cite{GLiSE2025} and its composition and size are presented in Table~\ref{tab:labels}.




\subsection{Embedding Model Selection}
Several embedding models were initially considered. We were particularly interested in Qwen3 \cite{qwen3embedding} and Gemini \cite{Google_GeminiEmbeddings2025} embedding models due to their strong performance on the MTEB benchmark \cite{MTEB_Leaderboard2025}. However, these options were ruled out because the Qwen3 models were too memory-demanding for regular personal machines, and because Gemini's rate limits were too strict and using it would require users to provide an additional API key.

Therefore, we opted for OpenAI's \textit{text-embedding-3-small} and \textit{text-embedding-3-large} models \cite{OpenAI_EmbeddingsGuide2025}. Since the tool already depends on their LLMs for query generation, adopting these models avoids the need for an additional API key, and their rate limits are sufficient for our needs.

\subsection{Model Training and Selection}

To identify an effective methodology to leverage embeddings to filter search results with our available resources, we explored and evaluated several alternatives.


\subsubsection{Filtering Models Inputs}
We experimented with different ways of transforming embeddings into model inputs. For each relevant textual field associated with retrieved results, we computed its embedding and paired it with the embedding of the search intent. For each such pair, we derived multiple features which we concatenated into input vectors for our ML models. We explored the following inputs combinations : (i) Cosine Distances, (ii) Euclidean Distances, (iii) L1 Distances, (iv) Cosine Distances and Euclidean Distances combined, (v) All three distances combined, (vi) Element-wise absolute difference vectors, (vii) Element-wise product vectors, and (viii) All features combined.

\subsubsection{Embedding dimensionalities}
We evaluated three embedding dimensionalities: 512, 1024, and 1536.

\subsubsection{Candidate models}

For each input representation, we trained the following classifiers: Gaussian Naive Bayes, Logistic Regression, XGBoost, LinearSVC, and Ridge \cite{scikit-learn, Chen_2016, sklearn_GaussianNB, sklearn_LinearSVC, sklearn_LogisticRegression, sklearn_RidgeClassifier}.

\subsubsection{Selection protocol}

All possible combinations of input representation, embedding dimensionality, and classifier were tested. We used a 50/50 train-test split. Hyperparameters were tuned on the training split using GridSearchCV \cite{scikit-learn_GridSearchCV}. We report balanced accuracy precision, recall, and F1 score for each trained model to assess filtering performance.

This process was repeated independently for each of the four result sources. For every source and embedding mode ("text-embedding-3-small" and "text-embedding-3-large"), we selected the best-performing configuration and integrated it into \textsc{GLiSE}.

\subsubsection{LLM baseline}

We also evaluated an LLM-based baseline (gpt-4o-2024-08-06) at temperature 0, using the most probable token for binary relevance prediction. We chose not to adopt this approach as it is slower, more costly, and inferior according to our baseline.

\subsubsection{Final selection}

The models ultimately integrated into \textsc{GLiSE} were those that achieved the best performance for each source. Their performance is reported in Table \ref{tab:mc}.
\begin{table*}[t]
\centering
\caption{Selection of relevance classification models by web source, embedding configuration, and classifier type. 
Reported metrics correspond to the best-performing setup.}
\label{tab:mc}
\footnotesize
\begin{tabular}{l l c c c c r}
\toprule
\textbf{Source} & \textbf{Representation} & \textbf{Balanced Accuracy} & \textbf{Precision} &  \textbf{Recall}& \textbf{F1 Score}  & \textbf{Best ML Model/Dimensions/Features} \\
\midrule
\multirow{3}{*}{Github Repos}
 & Text-Embedding-3-Small & 0.65 & 0.60 & 0.46 & 0.52 & GaussianNB / 1536 / all distances \\
 & Text-Embedding-3-Large & 0.71 &0.61  &0.64  &0.63  & XGBoost / 1024 / l1 distance\\
 & gpt-4o-2024-08-06 & 0.73 &0.71  &0.58  &0.64  & -\\
\hline
\multirow{3}{*}{Github Issues}
 & Text-Embedding-3-Small &0.72  &073  &0.55  &0.63  & GaussianNB / 1024 / all distances \\
 & Text-Embedding-3-Large &0.73  &0.69  &0.63  &0.66  & GaussianNB / 1536 / all distances  \\
 & gpt-4o-2024-08-06 & 0.61 &0.87  &0.24  &0.38  & -\\
\hline
\multirow{3}{*}{StackOverflow}
 & Text-Embedding-3-Small &0.76  &0.91  &0.65  &0.76  & GaussianNB / 512 / overlap product vectors \\
 & Text-Embedding-3-Large &0.76  &0.88  & 0.71 &0.78  & Ridge / 1024 / Element-wise absolute difference  \\
 & gpt-4o-2024-08-06 & 0.63 &0.79  &0.56  &0.66  & -\\
\hline
\multirow{3}{*}{Google}
 & Text-Embedding-3-Small &0.79  &0.76  &0.81  &0.78  & GaussianNB / 512 / Element-wise absolute difference  \\
 & Text-Embedding-3-Large & 0.80 &0.77  &0.82  &0.79  & GaussianNB / 1536 / Element-wise absolute difference  \\
 & gpt-4o-2024-08-06 & 0.71 &0.94  &0.45  &0.60  & -\\
\bottomrule
\end{tabular}
\end{table*}

\section{Usability Study}

To evaluate \textsc{GLiSE}'s user experience and determine whether it effectively reduces screening effort, we conducted an empirical study of its usability and usefulness with software-engineering professionals and researchers.

\subsection{Methodology}

Each participant was tasked with completing two comparable research objectives (different topics) in counterbalanced order: one \emph{manually} (baseline) and one using \textsc{GLiSE}. For each objective, participants had to identify $k$ relevant items (e.g., $k{=}10$) from the suggested sources.
After completing both objectives, participants were asked to answer questions in a dedicated Google Form.

To measure and compare participants' performance across both tasks, we used the following \emph{objective measures}: (i) \textbf{Time-To-First-Relevant (TTFR)}; (ii) \textbf{Items-To-10 (IT@10)} (the number of inspected items to find 10 relevant ones); (iii) \textbf{Screening/Research time} to reach 10 relevant; and (iv) \textbf{Coverage across sources} (distinct providers represented in the final set).



To complete our assessment of \textsc{GLiSE}'s user experience, we also administered the \textbf{System Usability Scale (SUS)} (10 items, 5-point Likert) \cite{Brooke1996SUS}, along with two single-item scales: \emph{Perceived Usefulness} and \emph{Intention to Reuse} (7-point). 

In total, the Google Form included five parts: (i) consent, (ii) demographic questions, (iii) objective measures, (iv) user experience assessment, and (v) fields for open-ended feedback.

We aggregated the results for Manual and \textsc{GLiSE}-assisted tasks using paired data. SUS was scored using the standard approch, where positive items are adjusted as \textbf{$response - 1$} and negative items as \textbf{$5 - response$}, the sum of the ten adjusted scores is then multiplied by $2.5$ to obtain a global SuS value between 0 and 100.

All study materials are provided in the replication package \cite{GLiSE2025}.

\subsection{Results and Analysis}



Five participants (four engineers and one researcher) completed both conditions (Manual and \textsc{GLiSE}-assisted). On average, the group had 2.6 years of in software-engineering experience and a moderate familiarity with evidence reviews (2.6 on a five point scale).


Participants were faster and more efficient with \textsc{GLiSE} than in the manual condition. Mean Time To First Relevant decreased from 158 seconds to 96 seconds which corresponds to a reduction of about 39 percent. Total screening time to reach ten relevant entries decreased from 20.0 minutes to 2.5 minutes, reflecting the consolidation of retrieval into a single prompt and the filtering provided by an accurate ML classifier (see Table \ref{tab:mc}).

\begin{table}[h]
\centering
\footnotesize
\setlength{\tabcolsep}{8pt}
\renewcommand{\arraystretch}{0.97}
\caption{Usability outcomes. Manual compared to \textsc{GLiSE}.}
\label{tab:utility}
\begin{tabular}{lrr}
\toprule
\textbf{Metric} & \textbf{Manual} & \textbf{\textsc{GLiSE}} \\
\midrule
TTFR (s) & 158 & 96 \\
IT@10 (items) & 24.8 & 1 \\
Time to 10 (min) & 20.0 & 2.5 \\
\midrule
SUS (0 to 100) & N/A & 81.0 \\
Usefulness (1 to 7) & N/A & 6.0 \\
Intention to reuse (1 to 7) & N/A & 6.0 \\
\bottomrule
\end{tabular}
\end{table}

Subjective feedback was consistent with the quantitative results. The mean SUS score for \textsc{GLiSE} was 81.0 with a sample standard deviation of 7.6. This value exceeds the commonly cited threshold of 68 for good usability \cite{LewisSauro2018_ItemBenchmarksSUS} and approaches the range often interpreted as excellent above 80. Perceived usefulness averaged 6.0 out of 7, and intention to reuse also averaged 6.0 out of 7. In open ended comments, participants appreciated using a dedicated tool to extract relevant grey literature for software engineering. They specifically valued the integration of GitHub and Stack Overflow within a single workflow which consolidated sources and shortened early screening. They suggested a leaner interface with lower information density fewer simultaneous elements and clearer progressive disclosure to limit cognitive load. Despite these requests, the overall assessment remained positive and participants indicated that they would adopt the tool in future evidence gathering.

\section{Threats to Validity}


\textit{Construct validity:} The concept of relevance is subjective, as two human evaluators may disagree on the relevance of specific results. In addition, class imbalance can negatively affect model training and evaluation. To mitigate these threats, we trained models with both uniform and balanced class weights, and \textsc{GLiSE} allows users to access results predicted as irrelevant if desired.

\textit{Internal validity:} To mitigate risks of data leakage and overfitting while training our classifiers, we performed deduplication and shuffled the data before splitting, and used per-source partitions for training and evaluation. We ensured reproducibility by fixing random seeds. In the usability study, we also randomized task order and provided a short practice task to participants. Additionally, the reported classifier performance was used for model selection and should not be interpreted as an unbiased evaluation.

\textit{External validity:} Results found with the tool may vary with time, location, or due to personalization. We timestamp versions, and release  the prompts and code to aid replication. We mitigate the effects of the small usability study sample through a counterbalanced design, objective measures, and SUS instrumentation.

\textit{Conclusion validity:} Our dataset is limited in size, which may affect our classifier's performance. To mitigate this threat, we varied both feature complexity (from compact cosine-based features to element-wise difference and product vectors) and model capacity, including simple classifiers such as Gaussian Naive Bayes.





\section{Conclusion}

In this paper, we presented \textsc{GLiSE}, a tool that automates grey literature retrieval and screening across multiple software engineering sources. Given a free-text prompt, \textsc{GLiSE} generates source-specific queries, gathers results, and filters and ranks them using embedding-based machine learning classifiers. Our usability study shows that participants clearly appreciated using \textsc{GLiSE} and were able to identify relevant sources in significantly less time. Looking ahead, we plan to extend \textsc{GLiSE} by integrating additional data sources, deploying a web-based version for accessibility, enlarging our dataset to improve our classifiers, and incorporate mechanisms to estimate the trustworthiness of retrieved items.


\bibliographystyle{ACM-Reference-Format}
\bibliography{bibli}

\appendix

\end{document}